\documentclass[aps,reprint]{revtex4-1}
\usepackage{graphicx}
\usepackage{amsmath}

\begin{document}%
%%%%%
%
% Titel
%
%%%%%
\title{On the Classical Description of Nuclear Motion}
\date{\today}
\author{Irmgard Frank}
\homepage{http://www.theochem.uni-hannover.de/}
\email{irmgard.frank@theochem.uni-hannover.de}
\affiliation{Theoretische Chemie\\ Leibniz Universit\"a{}t Hannover}
%%%%%
%
%
%
%%%%%

\begin{abstract}

Severe methodological and numerical problems of the traditional quantum mechanical approach to the description of
molecular systems are outlined. To overcome these, a simple alternative to the Born-Oppenheimer approximation
is presented on the basis of taking the nuclei as classical particles.

Keywords: 
Nuclear motion, Born-Oppenheimer approximation, molecular dynamics
\end{abstract}

\maketitle

\section{Introduction}

Essentially all quantum chemical calculations are based on the Born-Oppenheimer approximation.
As a starting point, one writes down the Schr{\"o}dinger equation for both the electrons and the nuclei:

\begin{equation} \label{1}
\hat{H}_{total} \Psi(\{{\bf r}_i\},\{{\bf R}_I\}) = E_{total} \Psi(\{{\bf r}_i\},\{{\bf R}_I\})
\end{equation}

with

\begin{equation} \label{2}
\hat{H}_{total}=\hat{T}_{nuc}+\hat{T}_{el}+\hat{V}_{nuc-nuc}+\hat{V}_{nuc-el}+\hat{V}_{el-el}
\end{equation}

Then one assumes that, due to the much higher masses of the nuclei, electronic and nuclear motion
can be separated. Introducing the product

\begin{equation} \label{3}
\Psi(\{{\bf r}_i\},\{{\bf R}_I\})=\Psi_{el}(\{{\bf r}_i\};\{{\bf R}_I\})\Psi_{nuc}(\{{\bf R}_I\})
\end{equation}

one arrives within the Born-Oppenheimer approximation at the electronic Hamilton operator

\begin{equation} \label{4}
\hat{H}_{el}=\hat{T}_{el}+\hat{V}_{nuc-el}+\hat{V}_{el-el}
\end{equation}

and the electronic Schr{\"o}dinger equation

\begin{equation} \label{5}
\hat{H}_{el} \Psi_{el}(\{{\bf r}_i\}) = E_{el} \Psi_{el}(\{{\bf r}_i\})
\end{equation}

which is the basis for
standard quantum chemistry \cite{Levine}. As such, it is one of
the best tested physical equations. The accurate description of the molecular structure,
reaction energies, excitation energies and many other properties represents an enormous success of the theory. 
For obtaining correct potential energy curves from quantum chemical calculations, the interaction between the nuclei has to be added:

\begin{equation} \label{6}
E_{qc} = E_{el} + \sum_I\sum_{J>I}\frac{Z_A Z_B e^2}{4 \pi \epsilon_0 R_{AB}}
\end{equation}
\\

However, what is tested to such a high degree of accuracy is the Schr{\"o}dinger equation for the
electrons \eqref{5}, not at all the starting point \eqref{1}. In fact, solving it
as a whole, or also solving the separated expressions for the electrons and the nuclei
iteratively till self-consistency, leads to no meaningful solutions. The nuclear
wave function feels the electronic cloud and is attracted by it, hence it diffuses.
The more diffuse nuclear wave function leads to a less localized electronic cloud
which gets more diffuse if it is computed again on the basis of the new nuclear
wave function. The result is a homogeneous gas for both the electrons and the nuclei.
There is no operator which could possibly extract any molecular information from
such a wave function (unless the operator
already contains the information for a certain molecule).
The obvious conclusion is that the starting point for the
Born-Oppenheimer approximation \eqref{1} is wrong, even if it is well established.
The nuclei must be described in a different way. \\

Actually, since the atom was separated in our theoretical description in a very asymmetric way 
in a localized nucleus and an electronic cloud, we had no real reason to assume that they are both described
by the same equation. But what alternatives do we have?
While \eqref{1} does not yield a meaningful result, molecular dynamics calculations
which treat the nuclei classically are surprisingly successful.
Are we allowed to think that the
classical treatment of the nuclei may be the way better starting point? What would this lead to? \\

The quantum mechanical treatment of the nuclei is used to explain for example vibrational spectra \cite{Levine}.
However, the most successful computations of vibrational spectra are done classically \cite{Silvestrelli1997,Vandevondele2012}
and it does not seem that quantum theory is needed in this field.
The same accounts to the branching into different reaction channels when simulating chemical
reactions. Classical chaos in a many-particle system is fully sufficient to lead to different reaction products
as could be shown for quite a few systems (e.g. \cite{Nonnenberg2008,Hofbauer2012}).
Actually, the good performance of density functional calculations in the framework of 
first-principles molecular dynamics for the description of chemical reactions
is traditionally attributed to error compensation as in these calculations the zero-point energy is neglected.
This zero-point energy is the most obvious difference between classical and quantum mechanics and
certainly the measure to which degree a motion is to be described quantum mechanically.
What is the experimental evidence for it? It is qualitatively known that organic reactions may run slower,
if hydrogen is replaced by deuterium (kinetic isotope effect). This is usually attributed to
a difference in the zero point energies. However, also classically changing the mass influences the motion
and one obtains slower motions for larger masses. What about the thermodynamic isotope effect?
The thermodynamic difference between the enthalpy for a certain reaction and the enthalpy for the
same reaction using deuterated compounds should be a most distinctive measure for quantum mechanical effects.
Unfortunately it is small for most reactions. A reaction where a clear difference of 2 to 3 kcal/mol could be seen is,
for example, N$_2$ + 3 H$_2$ $\to$ 2 NH$_3$, compared to N$_2$ + 3 D$_2$ $\to$ 2 ND$_3$. One would believe
that the corresponding data for such reactions are contained in the relevant data bases \cite{NIST,Chase1985}. However,
what is given there for the deuterated reaction is the reaction enthalpy as measured for the undeuterated
reaction plus the zero point energy as {\bf computed} from the vibrational spectra, already using the
assumption that the nuclear motion is purely quantum mechanical. \\

%There is an even more critical test: The comparison of the spectra of H$_2$ and D$_2$. 
%H$_2$ is excited to a weakly bound state. Assuming the validity of the Schr{\"o}dinger equation for the nuclei,
%one can estimate the difference in the ground and excited state zero point energies
%for H$_2$ and D$_2$ from the vibrational frequencies of the relevant states.
%For the lowest $^1\Sigma_u^+$ state this gives an energy difference between the
%H$_2$ and D$_2$ excitations of 890.70 cm$^{-1}$/2, for $^1\Pi_u$ 571.86 cm$^{-1}$/2, and for $^1\Sigma_g^+$
%481.23 cm$^{-1}$/2.
% 4401.21 - 1358.09 - 3115.50 + 963.08
% 4401.21- 2443.77 - 3115.50 + 1729.92
% 4401.21- 2588.9 - 3115.50 + 1784.42
%However, the data
%listed in the NIST data base for the lowest transitions show essentially negligible energy differences
%for the lowest transitions:
%$^1\Sigma_u^+$: H$_2$: 90203.3 cm$^{-1}$, D$_2$: 90633.79 cm$^{-1}$, difference: 430.49 cm$^{-1}$;
%$^1\Pi_u$: H$_2$: 99120.1 cm$^{-1}$, D$_2$: 99409.18  cm$^{-1}$, difference: 289.08 cm$^{-1}$
%$^1\Sigma_g^+$: H$_2$: 00082.3 cm$^{-1}$, D$_2$: 100128.1 cm$^{-1}$, difference: 45.8 cm$^{-1}$ 
%\cite{NIST,Dieke1936,Dieke1965,Dabrowski1974}.

%Actually, quantum chemists usually do not correct electronic spectra for zero point energies.
%Where ZPE calculations are typically used is to correct MP2 barriers which, however, may be too high
%for other reasons. \\

If we believe that there is no zero point energy then there is little room for a
quantum mechanical description of nuclear motion which is intrinsically tied
to uncertainty. We need an alternative approach.
In the present paper we investigate
the theory for a purely classical
description of nuclear motion, as a simple approach which avoids the most evident problems
with the basic theory, namely the diffusion of the nuclei. As such it is used nowadays in most
computations of molecular motion, but is viewed as an approximation to the equations following
from the Born-Oppenheimer approximation. Even if we change this view, a classical treatment is certainly not
the answer to all questions. It is obvious that such a description, just like the traditional
quantum chemical one (equation \eqref{1}), does not contain nuclear spin or phenomena like nuclear
fission, which the true, complete nuclear wave function should certainly describe. We are interested
in phenomena that occur on the scale of about 10$^{-11}$ m to 10$^{-10}$ m and do not describe
the components of the nucleus nor do we ever leave the reign of electrostatics. For the description
of most chemical phenomena this is sufficient and we are far from going beyond that. The picture we use
is: We observe that atoms move and we call the centers of this motion nuclei, assigned with masses
and charges. These centers do not diffuse nor split
by construction and a classical description of their motion is appropriate.
The rest of the atom is well 
described by the electronic Schr{\"o}dinger equation.

We start from \eqref{2} and apply it to the electronic wave function only. We do {\bf not} resort
to the Born-Oppenheimer approximation, but just assume that our nuclei are classical particles.
Note, that we also do not resort to the Ehrenfest theorem
which is derived from the time-dependent Schr{\"o}dinger equation.

\begin{eqnarray} \label{7}
\nonumber \hat{H}_{total} \Psi(\{{\bf r}_i\},\{{\bf p}_i\};\{{\bf R}_I\},\{{\bf P}_I\}) \\
%=\{\hat{T}_{nuc}+\hat{T}_{el}+\hat{V}_{nuc-nuc}+\hat{V}_{nuc-el}+\hat{V}_{el-el}\} \Psi(\{\bf{r}_i\},\{\bf{p}_i\};\{\bf{R}_I\},\{\bf{P}_I\}) \\
=E_{total} \Psi(\{{\bf r}_i\},\{{\bf p}_i\};\{{\bf R}_I\},\{{\bf P}_I\})
\end{eqnarray}

Since the wave function depends only parametrically on the nuclear positions ${\bf R}_I$ and momenta ${\bf P}_I$ (or velocities ${\bf V}_I$), one obtains immediately:

\begin{eqnarray} \label{8}
&\nonumber \left[\sum_I \frac{1}{2} M_IV_I^2 + \hat{H}_{el} + \sum_I\sum_{J>I} \frac{Z_I Z_J e^2}{4 \pi \epsilon_0 R_{IJ}}\right] \Psi(\{{\bf r}_i\},\{{\bf p}_i\}) \\
&= E_{AIMD} \Psi(\{{\bf r}_i\},\{{\bf p}_i\})
\end{eqnarray}

Starting again from \eqref{2} for
classical nuclei, one obtains the usual equations for nuclear motion (see textbooks of classical mechanics):

\begin{equation} \label{9}
M_I \frac{\partial^2}{\partial t^2}{\bf R}_I= \frac{\partial}{\partial {\bf R}_I} E_{qc}
\end{equation}

In \eqref{8} the energy is termed as E$_{AIMD}$, as it is the energy expression which is used in ab-initio
molecular dynamics calculations. It is most intuitive: the kinetic energy of the nuclei and the interaction between
the nuclei have to be added to the energy of the electronic Hamilton operator in order to achieve energy conservation.
In the limit of T$\to$0, that is vanishing nuclear velocities, E$_{AIMD}$ approaches E$_{qc}$ as computed from equation \eqref{6}.
Equations \eqref{8} and \eqref{9} describe what is -- rather inadequately -- referred to as Born-Oppenheimer 
molecular dynamics.
Using the time-dependent Schr{\"o}dinger equation instead of the time-independent Schr{\"o}dinger
equation leads to what is -- not more adequately -- called Ehrenfest dynamics.
Car-Parrinello molecular dynamics affords an additional assumption \cite{Marx2009}. Born-Oppenheimer and Car-Parrinello
molecular dynamics are well tested, mostly using the density functional approximation which
is then summarized as first-principles molecular dynamics. Way less is known about the performance of Ehrenfest dynamics
which is technically less simple.
It should be possible to describe
non-Born-Oppenheimer effects on this basis. Correctly maintaining energy conservation 
during an Ehrenfest dynamics should be sufficient to obtain these phenomena. Of course the
electronic wave function must be described flexibly enough to fully allow arbitrary changes of configurations.
Otherwise one is limited to either the diabatic or adiabatic pathways. This can be sufficient
to describe a certain reaction \cite{Frank2007}, but does not allow a general description of non-Born-Oppenheimer effects.
For this something like a CASSCF wavefunction is needed.
Fulfilling this requirement in an Ehrenfest dynamics
is hard enough, however, 
a quantum description of the nuclei is not necessary. If during a photoreaction a
molecule changes its configuration to go to a higher-lying state, this is just due to energy conservation during nuclear motion.
Kinetic energy is converted into potential energy
and ab-initio molecular dynamics trajectories should be sufficient to describe this phenomenon.
To check this numerically, further developments of the presently implemented approximations \cite{Tavernelli2006}
are necessary and it will take at least another decade till there is a complete picture how well multi-configuration
Ehrenfest dynamics performs.
Many phenomena are already well understood on the basis of ab-initio molecular dynamics simulations,
but on a longer time-scale 
the simulation of non-Born-Oppenheimer effects are the important test cases for a time-dependent theory which describes
the nuclei purely classically and the electronic wave function using the Schr{\"o}dinger equation.

\section{Conclusions}

We show that using the assumption that the nuclei in quantum chemical calculations are classical particles,
one arrives very directly at the equations used in ab-initio molecular dynamics calculations,
without resorting to the Born-Oppenheimer approximation. The great success of the classical
description of nuclear motion in AIMD simulations as well as fundamental problems with the conventional 
quantum mechanical theory for the nuclei induce the idea that the classical description is even superior.
This could be substantiated for example by calorimetry experiments accurately measuring the thermodynamic isotope effect.

%\section{Acknowledgements}

\bibliography{../../literature.bib}

\end{document}